\documentclass[aps,reprint,superscriptaddress,prl]{revtex4-1}
\usepackage[T1]{fontenc}       % Use modern font encodings

\usepackage{color}
\usepackage{amsmath, amsthm, amsfonts}    % need for subequations
\usepackage{amssymb}
\usepackage{bm}
\usepackage{comment}
\usepackage[normalem]{ulem}
\usepackage{hyperref}   % use for hypertext links
\usepackage[dvipsnames]{xcolor}
\hypersetup{
    colorlinks,
    linkcolor={blue!80!black},
    citecolor={blue!80!black},
    urlcolor={blue!80!black}
}
\usepackage{graphicx}   % need for figures
\usepackage{subfig}     % use for side-by-side figures

\usepackage[makeroom]{cancel}
\usepackage[normalem]{ulem}

\newcommand{\mob}{\boldsymbol{\mathcal{M}}}
\newcommand{\mobF}{\boldsymbol{\mathcal{M}}_{3D}}
\newcommand{\angstrom}{\mbox{\normalfont\AA}}

\begin{document}

%%%%%%%%%%%%%%%%%%%%%%%%%%%%%%%%%%%%  Head  %%%%%%%%%%%%%%%%%%%%%%%%%%%%%%%%%%%%

\title{Solvent hydrodynamics enhances the collective diffusion of membrane lipids}

  \author{S. Panzuela}\email{sergio.panzuela@uam.es}
\affiliation{Department of Theoretical Condensed Matter Physics,
  Universidad Aut\'onoma de Madrid, 28049 Madrid, Spain}

\author{R. Delgado-Buscalioni}\email{rafael.delgado@uam.es}
\affiliation{Department of Theoretical Condensed Matter Physics,
  Universidad Aut\'onoma de Madrid, 28049 Madrid, Spain}

\begin{abstract}
The collective  motion of membrane lipids
over hundred of nanometers  and  nanoseconds is essential for
the formation of submicron complexes of
lipids and proteins  in the cell membrane.
These dynamics are difficult   to   access   experimentally
and are currently poorly  understood.
One of the conclusions of the celebrated Saffman-Debr\"uck (SD) theory
is that lipid disturbances smaller than 
the Saffman length (microns) are not
affected by the hydrodynamics of the embedding solvent.
Using molecular   dynamics  and   coarse-grained   models  with   implicit hydrodynamics we show that this is not true. Hydrodynamic interactions between the membrane and the
solvent strongly enhance the short-time
collective diffusion of lipids at all scales.
The momentum transferred between the membrane  and the  solvent
in normal direction (not considered by the SD theory)
propagates tangentially over  the membrane
inducing  long-ranged repulsive  forces amongst  lipids.
As a consequence the lipid collective   diffusion coefficient increases proportionally to the disturbance wavelength. We find quantitative agreement
with the predicted anomalous diffusion in quasi-two-dimensional
dynamics, observed in  colloids confined to a plane but
embedded in 3D solvent. 
\end{abstract}

\date{today}
\maketitle

The  study  of  lipids,  as   building  units  of  the  cell
membranes,  has   been  intensive   since  their  discovery   in  1925
\cite{Gorter1925}.   While  the  membrane equilibrium  structures  are
relatively well understood, the collective dynamics of lipids is still
relatively unexplored  \cite{ilpo2013,rheinstadter2008prl,2016Zher}.
The strong coherence between lipid displacements observed from few  to hundred
nanometers over less or about hundred of nanoseconds \cite{ilpo2009faraday,ilpo2013,rheinstadter2008prl}
is essential to the membrane fluidity and is crucial to biological functions such as
the formation  of nanometric pores \cite{2016Zher}, the kinetics of
submicron  complexes (lipid    rafts) \cite{ilpo2009faraday},
protein transport, transduction or gating mechanisms  \cite{carquin2016}.
This mesoscopic spatio-temporal scale is difficult to access experimentally
\cite{rheinstadter2008prl,ilpo2013,tobias2010jacs,tabony1991biochimica,ilpo2009faraday,srivastava2012hybrid,armstrong2013hindawi}, being too large for neutron scattering or spin  echo \cite{armstrong2013hindawi} but
too fast and small for standard fluorescence labeling techniques \cite{2010Machan}.
At the nanometric border           of          this           ``mesoscale          gap'' \cite{armstrong2013hindawi}
quasi-elastic      neutron      scattering     (QENS)      experiments
\cite{tobias2008jcp,tobias2010jacs,armstrong2011softmatter}       have
recently  measured two  relaxation mechanisms:  one compatible  with a
fast,  purely diffusive  lipid motion  and a  slower, ballistic  mode,
which  was interpreted  as nanometric  currents of  lipids propagating
with velocities much smaller  than the thermal value $(k_BT/m)^{1/2}$.
Molecular  simulations  \cite{falck2008jacs,ilpo2013}  are  consistent
with this  view, which implies  that lipids diffuse in  a coordinated,
hydrodynamic-like, fashion instead than by discrete ``jumps'' out from
molecular cages. Coarse grained (CG) models
with two-dimensional (2D) hydrodynamics find that the correlations
between lipid displacements \cite{ilpo2013} span  over more than 10 nm
and microseconds \cite{ilpo2013,ilpo2009faraday}.  Correlations beyond
10     nm    were     also     inferred     by    QENS     experiments
\cite{rheinstadter2008prl}.

Since        1975,   the  elegant     Saffman-Delbr{\"u}ck        theory
\cite{saffman1975pnas,frank2010}  has   been  used  to   describe  the
interaction between the membrane and solvent hydrodynamics.  Its great
merit is to show that the solvent's {\em tangential} friction slows down the
collective motion of lipid flow patterns if their wavelength $\lambda$
is  larger  than  a  certain   cut-off  distance:  the  Saffman  length
$\lambda_S$.     This length    $\lambda_S=\eta_m/(2\eta)$
(proportional  to the  ratio  between the  surface  viscosity and  the
solvent viscosity) is typically in  the micron scale. Therefore,
it is currently thought that, at submicron  scales,
the momentum-exchanged with the solvent is negligible
so lipid dynamics approximately conserves the in-plane momentum, as in 2D hydrodynamics.
For these scales, all-atom   molecular  dynamics  (MD)   \cite{2013Bennet} are prohibitive,
while continuum fluid dynamics \cite{frank2010}
does not resolve lipid motions. Therefore, different CG models
have been used to study collective motions in the mesoscale domain \cite{ilpo2013,ilpo2009faraday,marrink2013rsc,2004Voth}.
Albeit, these works focused on momentum-conserving, 2D hydrodynamics.

In this Letter we use Martini-MD  simulations \cite{marrink2013rsc}
and CG  models equipped with an immersed boundary description  of the
3D-hydrodynamics  \cite{fluam-stokes} to show  that  the  solvent
hydrodynamics strongly enhances the  collective diffusion of lipids at
all scales,  even below  $\lambda_S$. Notably, the  momentum exchanged
with  the solvent  in  the  {\em normal  direction}  spreads over  the
membrane acting like a repulsive hydrodynamic force between lipids. As
a consequence, the collective  diffusion increases without bounds with
the disturbance  wavelength.  This  type of {\em  anomalous collective
diffusion}   is   in   quantitative  agreement   with   the   quasi-2D
hydrodynamics observed in colloids confined  in a plane and surrounded
by   solvent   \cite{Lin1995pre},   recently   analyzed   in   several
works  \cite{bleibel2014softmatter,bleibel2017pre,panzuela2017pre}. In
membranes,  this  phenomena  introduces   a  so  far  unexplored  {\em
intermediate   dynamic  regime}   which  significantly   enhances  the
collective  diffusion at {\em short times} (over about $100 \mathrm{ns}$)
and affects a wide range of scales, from  nanometers to  microns.

%More generally, following Ref. \cite{2009Granick},
%this dynamics can be described as ``anomalous yet Brownian'':
%while the lipid self-diffusion is Brownian \cite{panzuela2017pre},
%The anomalous collective diffusion is activated by lipid density fluctuations 
%whose wavelength is larger than the diffusing elements \cite{bleibel2014softmatter,fluam2d}

\begin{figure*}[th]
  \includegraphics[width=0.6\linewidth]{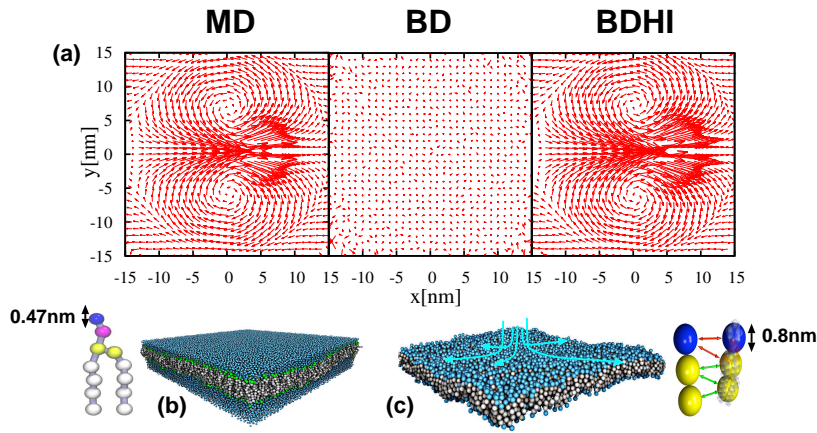}
  \caption{(a) Averaged velocity field
    evaluated from lipid displacements $\delta {\bf r}_2 ={\bf r}_2(\Delta t)-{\bf r}_2(0)$ with respect to
    a central tagged lipid
    at $\Delta t=400\mathrm{ps}$. 
    (b) and (c) are pictorial illustrations  of  the models taken from simulations.
(b), MD corresponds  to the  MARTINI model  for DPPC  with explicit  water
    \cite{marrink2013rsc}  and  BD  and   BDHI  simulations (c) to  the
    Cooke-Deserno model  \cite{cooke2005jcp}. The arrows drawn in (c) sketches normal
    momentum from the solvent spreading over the membrane.
    \label{fig:velfield}
  }
\end{figure*}

%Self and collective diffusion}
{\em Models with hydrodynamics}.
We solve the dynamics of membranes in periodic cubic boxes
of sizes up to $L\approx 88\mathrm{nm}$ using three different computational techniques
and two membrane models (more details in Supplementary Information, SI).
Both lipid models are depicted in Fig. \ref{fig:velfield}.
Simulations with greater resolution detail correspond to molecular dynamics (MD)
using the Martini force-field \cite{Martini} and solved by the GROMACS package \cite{Gromacs1,Gromacs2}.
Following the Martini description \citep{marrink2004coarse}, water molecules are explicitly resolved
while the membrane is formed by Dipalmitoylphosphatidylcholine (DPPC) lipids.
Simulations were performed at temperauture $T=310\mathrm{K}$.
The coarser membrane description, hereafter referred to as CG model,
is based  on the  ``dry membrane'' model  by Cooke and Deserno  \cite{cooke2005jcp} which
implements lateral  lipid-lipid interactions to stabilize  the bilayer
without  explicit water. The CG membrane matches the
compressibility and dominant peak in the static structure factor
of the MD (Martini) model (see SI).
Two different  dynamics were  implemented in the GC simulations.   On one
hand,  we  used the standard pure  Brownian  dynamics  (BD) method,
whereby  lipids  random displacements are completely uncorrelated.
The second approach is based on an {\em  implicit hydrodynamic  solvent}
solved by the  FLUAM package  \cite{fluam}. It basicly consists on
an immersed  boundary description in fluctuating hydrodynamics \cite{fluam1,fluam2,fluam-stokes}. Lipid beads are treated as immersed
particles and exchange momentum with the solvent's
fluctuating momentum-field. We worked in the Stokesian limit \cite{fluam-stokes}
where momentum propagates instantaneously (no   fluid-inertia).
This description is equivalent to Brownian   hydrodynamics with  the
Rotne-Prager-Yamakawa mobility and we shall refer these simulations as
Brownian    dynamics   with  hydrodynamic interactions, BDHI.

These three models (MD, BD and BDHI) treat momentum transfers in quite different
ways. While the BD model provides a vanishing correlation between lipids displacements, BDHI conserves momentum and lead to strong in-plane correlations.
In BDHI, any force acting on a lipid is transferred
to the fluid and, like in  the  Saffman  model \cite{saffman1975pnas,frank2010},  the fluid-lipid coupling is determined by the  no-slip constraint (see \cite{fluam1,fluam2}). The MD  model also  conserves total momentum
but this is exclusively transferred by {\em intermolecular collisions}
(either lipid-water or lipid-lipid interactions).
A first comparison between models is shown in Fig. \ref{fig:velfield}
by plotting the relative lipid-velocity field around a tagged lipid.
As expected, the BD scheme shows no trace of correlations.
But notably, despite their differences,
MD and BDHI yield a remarkably  similar vortical pattern
of correlated motions spanning over more than $10 \mathrm{nm}$.
It seems that momentum conservation is what really matters
to match MD and BDHI models. At first glance,
the flow pattern of Fig. \ref{fig:velfield} is similar to that found
in prior DPD 2D-simulations  \cite{ilpo2009faraday,ilpo2013}, however 
relevant differences shall be soon highlighted.

\begin{figure}[t]
\includegraphics[width=0.65\linewidth]{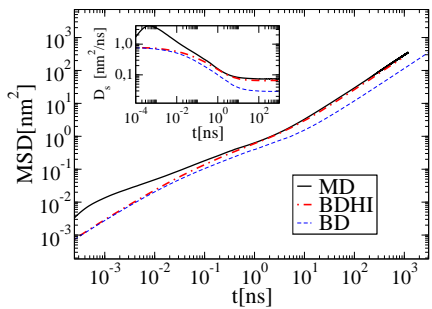}
  \caption{Mean square displacement of one lipid obtained from MD,
    BD and  BDHI models. In BD and  BDHI models the time
    is scaled to match the Stokes diffusion time $\sigma^2/D_0$,
    with $D_0=k_BT/(3\pi\eta \sigma)$ calculated using the viscosity of the
    Martini water-model \cite{Note1}.
    The  inset shows  the {\em  self} diffusion coefficient  $D_s(t)=\mathrm{MSD}(t)/(4t)$.
       \label{fig:dself}}
\end{figure}

{\em The lipid self diffusion} offers quantitative
comparison between models and experimental values.
Figure \ref{fig:dself} shows the lipid mean square displacement
of the lipid's head bead (see Fig. \ref{fig:velfield})
projected in the plane, $\mathrm{MSD}(t) = \langle ({\bf r}^{\parallel}_1(t)-{\bf  r}^{\parallel}_1(0))^2\rangle$
along with the {\em self} diffusion coefficient $D_s(t) =\mathrm{MSD}/(4t)$. All models present three dynamic regimes,
similar to those observed in membranes in the fluid-phase \cite{tobias2008jcp,armstrong2013hindawi}: a short-time diffusion
is followed by a sub-diffusive regime which finally leads
to a slower  long-time  diffusion  regime. The  intermediate
sub-diffusive   regime  approximately extends
from $0.01\mathrm{ns}$ to $10\,\mathrm{ns}$
and  is   characterized  by   $\mathrm{MSD}(t)\sim
t^\alpha$ with a sub-diffusive exponent $\alpha\approx  0.55$ 
quite    similar     in    all    models.
Fig. \ref{fig:dself} shows an excellent agreement between
MD and BDHI models (unit re-scaling explained in footnote \cite{Note1}
%\footnote{In terms of the
%Stokes diffusion coefficient $D_0=k_BT/(3\pi\eta \sigma)$
%we get $D_s^{(\ell)}=(0.086\pm 0.002)\,D_0$ for BDHI.
%To compare with MD we use $D_0\simeq  75 \mathrm{A}^2/\mathrm{ns}$ obtained
%from the viscosity of the Martini water model  $\eta=7\times 10^{-4}\mathrm{Pa\cdot  s}$ at $T\approx  300\mathrm{K}$ and the CG lipid-head  diameter $\sigma=0.8\mathrm{nm}$.}
), with long-time diffusion coefficients $D_s^{(\ell)}=7.8 \angstrom^2/\mathrm{ns}$ (MD)
and $6.4 \angstrom^2/\mathrm{ns}$ (BDHI) consistent with
QENS experiments (e.g. Ref. \cite{armstrong2011softmatter,armstrong2013hindawi} reports $6.3 \angstrom^2/\mathrm{ns}$). 
A comparison between BD and BDHI  indicates that
hydrodynamics increase the self-diffusion (for BD,
$D_s^{(\ell)}=2.9 \angstrom^2/\mathrm{ns}$).
This also happens in  dense colloids
due to  the reflection  of the  scattered
momentum,  out  and  back  to  each particle \cite{Dhont-book}.

\begin{figure}[!h]
  \includegraphics[width=0.6\linewidth]{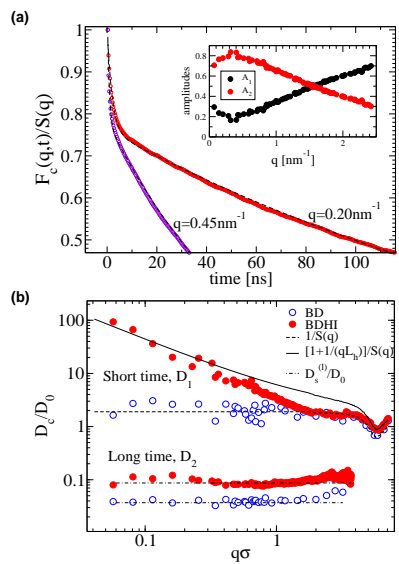}
  \caption{(a)   Time  dependence   of  the   normalized  intermediate
    scattering function of  a BDHI simulation. Dashed   lines    are   doubly   exponential   fits
    in Eq. \ref{ffit}.
     The  amplitudes of the fast  ($A_1$) and
    slow ($A_2$) relaxations are shown in the inset.(b) Short-time and
    long-time  collective  diffusion   coefficients  $D_1$  and  $D_2$
    extracted from  Eq. \ref{ffit}  and  plotted  against 
    $q\sigma$  with $\sigma=0.8\mathrm{nm}$.
    Solid and  dashed  lines following short-time relaxation data $D_1$
    are the theoretical predictions for BDHI and BD simulations (see text).
 At long-times, $D_2$ coincides with the
     long-time {\em self} diffusion (dot-dashed lines).
    \label{fig:fcol}
  }
\end{figure}

{\em Anomalous collective diffusion.}
The significant correlations between lipid displacements
observed in Fig. \ref{fig:velfield}
indicate  strong collective effects.
A way to analyze these correlations is provided by
the {\em collective}  intermediate scattering function $F_{c}(q,t)$
obtained  from the time-evolution of
the Fourier transform of the spatial correlations
in lipids density,
\begin{equation}
F_c(q,t)=\left<\rho\left(\bold{q},t\right)\rho\left(\bold{q}^*,0\right)\right>,
\label{fcol}
\end{equation}
where ${\bf q}$ is a wavevector in the membrane plane and we
assume that the system is isotropic in the plane, so that $F_c(q,t)=F_c({\bf q},t)$.
Figure \ref{fig:fcol}(a) shows exemplary results for $F_c(q,t)$.
We find that $F_c(q,t)$ can be fitted using a two-relaxation model,
\begin{equation}
\frac{F_c(q,t)}{S(q)} \approx A_1(q) \exp[- t/\tau_1] +A_2(q) \exp[ -t/\tau_2],
\label{ffit}
\end{equation}
where $S(q)=F_c(q,0)$ is the static structure factor.
The relaxation times $\tau^{-1}_1 \equiv D_1(q) q^2$ and $\tau^{-1}_2\equiv D_2(q) q^2$ introduce two
effective  diffusion coefficients ($D_1>D_2$) 
related to  short-time  and  long-time relaxation  mechanisms.
A similar two-relaxation model has been used
in QENS experiments to fit $F_c$  \cite{armstrong2013hindawi} at $q= 0.5 \angstrom^{-1}$
and the scattering spectra (see Supporting Information of Ref. \cite{tobias2010jacs}).
For     $q=0.5\angstrom^{-1}$     Armstrong     {\em    et     al.}     find
$\tau_1=1/(D_1q^2)=0.05\mathrm{ns}$ and $\tau_2=1/(D_2q^2)=0.9\mathrm{ns}$ which  provide $D_1=80 \angstrom^2/\mathrm{ns}$  and  $D_2=   4.5\angstrom^2/\mathrm{ns}$. At  the equivalent  wavenumber $q\sigma  \simeq 4$
(with $\sigma=8  \angstrom$), BDHI simulations yield respectively
$D_1 \approx D_s^{(s)} = 75 \angstrom^2/\mathrm{ns}$ and
$D_2 \approx D_s^{(\ell)} =6.4 \angstrom^2/\mathrm{ns}$
in excellent agreement with these experiments.
Our CG correctly capture the collective dynamics
at molecular wavelengths and now we analyze larger scales.

\begin{figure}[hbtp]
  \includegraphics[width=0.6\linewidth]{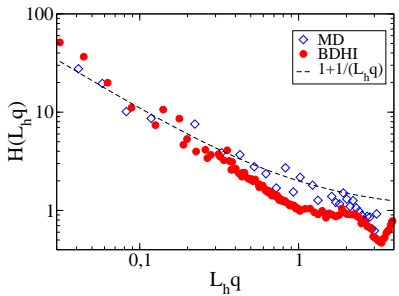}
  \caption{Hydrodynamic  function  $H(q)=D_1(q) S(q)/D_0$  versus  the
    non-dimensional   scale  $qL_h$.    Solid   line  represents   the
    theoretical  relation for  q2D dynamics  $H(q)=1+(qL_h)^{-1}$ with
    $L_h\simeq  0.53  \,\mathrm{nm}$  (MD)  and  $0.44\,  \mathrm{nm}$
    (BDHI).
    %For BDHI we  used  $D_0=k_BT/(3\pi\eta  \sigma)  \simeq  58\,\angstrom^2/\mathrm{ns}$
    For  MD we  used $D_0=4\mathrm{nm}^2/\mathrm{ns}$
    corresponding to the peak of  $D_s(t)$ at short-time (see inset of Fig. \ref{fig:dself}).
%    Also, $D_0= 40\angstrom^2/\mathrm{ns}$ approximately correspond to the Stokesian diffusion of a oligo-chain (lipid) with 10 beads of diameter $\sigma=0.47\mathrm{nm}$.
    \label{fig:h1}
    }

\end{figure}

%\begin{figure}[t]
%  \centering
%  \includegraphics[width=1\linewidth]{./collective_diffusion_comparative.eps}
%  \caption{Collective diffusion coefficients $D_1$ and $D_2$ related to
%  short time and long time relaxation (see Eq. \eqref{ffit}) scaled with the
%  (reference) free-diffusion coefficient $D_0$ and  plotted against  $q\sigma$
%  where $\sigma=0.8\mathrm{nm}$ is the lipid diameter.
%  Solid and dashed lines provide the theoretical expectations for each
%  case (with and without solvent hydrodynamics) at short and long times.
%  Results corresponds to the Cooke-Deserno membrane model \cite{cooke2005jcp}
%  with (BDHI) and without (BD) hydrodynamics.
%    \label{fig:d1d2} }
%\end{figure}

The values of $D_1$ and $D_2$ for the BD and BDHI models are plotted
in Fig.  \ref{fig:fcol}(b) against $q\sigma$.  We start by analyzing
the short-time collective diffusion, $D_1$.  At molecular wavelengths
$q\sigma >2$ all models yield a qualitatively similar trend
$D_1(q) \approx D_0/S(q)$.  In MD, we find that $D_0\approx
4\mathrm{nm}/\mathrm{ns}$ agrees with the peak of $D_s^{(s)}$ at short
times (inset of Fig. \ref{fig:dself}).  The dynamics of larger density
fluctuations ($q\sigma<1$) is however strongly modified by
hydrodynamic correlations.  Figure \ref{fig:fcol}(b) reveals an
anomalous increase of the short-time diffusion as $q$ decreases.  The
trend we observe $D_1 \sim 1/q$ is strongly reminiscent of the anomalous
collective diffusion in colloids confined to move in a plane
but embedded in 3D solvent, a set-up which usually called
quasi-two-dimensional (q2D) \cite{bleibel2017pre,panzuela2017pre}.
Under hydrodynamic interactions the short-time diffusion
is expressed as $D_1=[H(q)/S(q)]\,D_0$, where $H(q)$
is called the hydrodynamic function \cite{Dhont-book}.
Notably, under q2D dynamics, it has been
proved \cite{bleibel2014softmatter} that $H(q)$ diverges like
$H(q)=1+(qL_h)^{-1} + O(q)$. The anomalous diffusion
is already felt at molecular scales because
the ``hydrodynamic length'' $L_h=\sigma/(3\phi)$
is of molecular size ($L_h=0.44\mathrm{nm}$ for
BDHI and $0.53\mathrm{nm}$ for MD, calculated using the lipid head
surface fraction $\phi\simeq 0.6 \angstrom^{-2}$ and $0.3 \angstrom^{-2}$, respectively).
Figure \ref{fig:fcol}(b) shows
that the q2D theory agrees extremely well with our results for $D_1$
for $q\sigma > 0.4$.  Deviations at larger $q$ are expected because
the theory \cite{bleibel2014softmatter} only considers the Oseen
contribution of solvent's mobility (only valid at long distances, see
e.g. Ref.  \cite{panzuela2017pre} for a study with purely repulsive
colloids).

In view  of the different ways BDHI and MD  models carry out momentum transfer, we
need to  validate this phenomena  against the more detailed  MD model.
The  hydrodynamic function  $H_1(q)  = D_1(q)  S(q)  /D_0$ permits  to
compare  MD and  BDHI models  in  one single  master-curve, which is
shown  in Fig. \ref{fig:h1}.  The agreement  is extremely  good and  both models agree  remarkably  well with  the  q2D  theory (i.e.,  quantitatively, without fitting parameters).
This central result shows that hydrodynamic  interactions  between  the membrane and the  ambient fluid leads to a  significant enhancement of collective lipid  diffusion compatible  to that observed  in colloidal q2D  dynamics \cite{bleibel2014softmatter}.   While  in colloids,  the confinement arise from an external force field \cite{panzuela2017pre}, in membranes, it arises from the internal elastic forces
acting in normal direction to the plane.  These forces transfer normal
momentum  to the  surrounding fluid  which spreads  {\em tangentially}
over  the membrane. This mechanism is sketched with with arrows in  Fig.  \ref{fig:fcol}(c).
The resulting  collective drag  acts  like a  long-ranged repulsive
force ${\bf f}_{q2D}$ between  lipids. Technically, this force is  a form of {\em thermal drift}
\cite{Risken-book} arising  in the  presence of a mobility gradient  ${\bf f}_{q2D}=k_BT \nabla \cdot \mob$.
In q2D, the resulting current is proportional to the (lipid) density fluctuations \cite{Pelaez2018},
being controlled by the Oseen's contribution of the {\em solvent's} mobility $\mobF$.
The  solvent  is  incompressible  so  $\nabla\cdot  \mobF=0$,  but
a non-zero  divergence  appears when evaluated  {\em  in  the  plane} ($z=0$ and ${\bf  r}={\bf  r}^{\parallel}$) 
simply   because  $\nabla_{{\bf  r}^{\parallel}}\cdot \mobF  = -\partial_z \mobF \ne  0$. Particles
confined  to  move in  the  plane  are  thus  exposed to  a  repulsive
hydrodynamic force  ${\bf f}_{q2D}= k_BT  \nabla_{{\bf r}^{\parallel}}
\cdot   \mobF({\bf   r}^{\parallel},z=0)$   which   is   long   ranged
$f_{q2D}\propto    1/r^2$     because    $\mob_{3D}     \propto    1/r$
\cite{bleibel2016jp,panzuela2017pre,Pelaez2018}.

As an important aside, we note that
the long time collective diffusion coefficient $D_2$,
reaches a roughly constant value independent  on $q$
[see Fig. \ref{fig:fcol}(b) for BD and BDHI models].
This means that at long times (about $100\mathrm{ns}$)
the dynamics will gradually recover its normal diffusive character.
Remarkably, in both models (BD and BDHI) $D_2$
coincides (within error bars) with the long-time
{\em self}  diffusion  coefficient $D_s^{(\ell)}$.
This result proves that the q2D anomalous diffusion
does not generally lead to a divergent long-time diffusion
($D_c^{\ell} \sim 1/q$) in contrast with what it has been
pointed out by some authors \cite{bleibel2017pre}.
Our MD simulations are not long enough to resolve the long-time regime (SI)
and we cannot confirm a similar outcome $D_c^{\ell} \rightarrow D_s^{\ell}$ in this case.

{\em Conclusions.} From the standpoint of membrane modeling, we show
that a coarse {\em dry} membrane model equipped with the fluctuating
immersed boundary method (FIB) \cite{fluam1,fluam2,fluam-stokes}
reproduces the correlations of lipid's displacements observed in
molecular representations with explicit solvent molecules.  This type
of CG model with implicit hydrodynamics will allow to 
explore a significantly larger range of the huge spatio-temporal scales present in membrane dynamics. In passing, we
note that FIB methods are also able to correctly describe the shape
fluctuations of a membrane \cite{wang2013dynamic}, however, to the
best of our knowledge they have not been so far used to study lipid
dynamics.

From the perspective of lipid membrane dynamics,
present findings indicate that below $100 \mathrm{nm}$ and $100 \mathrm{ns}$ the highly correlated pattern of lipid displacements
is mainly a consequence of hydrodynamic interactions
with the solvent. {\em Normal} (out-of-plane)
momentum, exchanged between the incompressible fluid and the membrane elastic forces, induces long-ranged repulsive hydrodynamic interactions
in the plane which strongly enhance the short-time collective
diffusion of lipids at small wavenumbers $q\sigma <1$.
Notably, this phenomena is not described by the Saffman theory \cite{saffman1975pnas,frank2010} which otherwise explains the slowing down
of large lipid disturbances (above one micrometer) due to the action of the solvent {\em tangential friction}. An unified view of the solvent-membrane
hydrodynamics, from nanometers to microns will require
taking this novel effect into consideration.

%The observed phenomena is fully consistent with the anomalous
%diffusion enhancement predicted for quasi-2D dynamics
%Correlations due to the propagation of
%lipid-lipid (momentum-conserving) interactions, start to take over
%only after about $100 \mathrm{ns}$ leading to the Saffman dynamics.
%We note however, that the box-sizes deployed in this work are not long enough to observe the Saffman regime
%and, for the scales sufficiently resolved, the long-time lipid mobility obtained from our simulations
%looks like true 2D hydrodynamics.
%previous DPD simulations below $\lambda_S$ \cite{ilpo2013}).

Finally, some remarks on the experimental verification of this new
phenomena are due. According to our observations, lipid's density
patterns will initially decorrelate over {\em short times} of order
$\tau_{s} \sim 1/(q^2 D_1)$.  For $q\sigma<0.1$
(i.e. $q<1.4\mathrm{nm}^{-1}$) the intermediate scattering function
drops a $20\%$, i.e. to $F_c(q,\tau_{s})/S(q)\simeq 0.8$ after
$t\simeq \tau_s \sim 10\mathrm{ns}$. The anomalous diffusion yields
$D_1\sim D_0/(qL_h)$ so the initial decorrelation propagates with a
characteristic {\em velocity} $v_{q2D}= q D_1 \simeq D_0/L_h \sim
2 \mathrm{nm}/\mathrm{ns}$.  Interestingly, this velocity is quite compatible with
the ballistic relaxation velocities measured in QENS experiments (see SI of Ref. \cite{tobias2010jacs}).
At larger scales, propagating  at speed $v_{q2D}$, the  anomalous diffusion should
lead to an appreciable decorrelation of micron-size fluctuations in about one micro-second. This gives an idea
of the  experimental difficulty  in measuring this  collective effect.
Experimental confirmation  of this fast, but  strong, transient regime
might come from spin-echo experiments at low  wavenumber \cite{tobias2010jacs,armstrong2011softmatter} or
from high  time  resolution fluorescence  correlation  spectra \cite{2010Machan}
performed in vesicles; a spherical geometry for which
the theory for q2D diffusion has been recently derived \cite{dominguez2018}.
The strong enhancement of collective diffusion might prove to be relevant for
the formation of nanopores \cite{2016Zher} or on the kinetics
of lipids and proteins complexes \cite{carquin2016,ilpo2009faraday}.

{\em Acknowledgments.}
  We thank  Aleksandar Donev for  discussions and critical  reading of
  the manuscript and Raul P\'erez for helping us with figures and coding. We
  acknowledge  the support  of  the Spanish  Ministry  of Science  and
  Innovation MINECO  (Spain) under grant FIS2013-47350-C5-1-R  and the
  ``Mar\'{\i}a de Maeztu''  Programme for Units of  Excellence in R\&D
  (MDM-2014-0377). Part  of the  simulations were done  in Marenostrum
  under grant FI-2017-2-0023. R.D-B acknowledges support the donors of
  The American  Chemical Society  Petroleum Research Fund  for partial
  support of this research via PRF-ACS ND9 grant.

\end{document}